\documentclass[12pt]{article}
\usepackage{jcappub}

\begin{document}

%\usepackage{hyperref}% add hypertext capabilities
%\usepackage[mathlines]{lineno}% Enable numbering of text and display math
%\linenumbers\relax % Commence numbering lines

%\usepackage[showframe,
%Uncomment any one of the following lines to test
%%scale=0.7, marginratio={1:1, 2:3}, ignoreall,
% default settings
%%text={7in,10in},centering,
%%margin=1.5in,\affiliation
%%total={6.5in,8.75in}, top=1.2in, left=0.9in, includefoot,
%%height=10in,a5paper,hmargin={3cm,0.8in},
%]{geometry}

\title{Secluded and Putative Flipped Dark Matter and Stueckelberg Extensions of the Standard Model}

\author[a]{E. C. F. S. Fortes}
\emailAdd{elaine@ift.unesp.br}

\author[b]{V. Pleitez}
\emailAdd{vicente@ift.unesp.br}

\author[c]{F. W. Stecker}
\emailAdd{floyd.w.stecker@nasa.gov}

\affiliation[a]{Universidade Federal do Pampa, Rua Luiz Joaquim de Sá Brito, s/n, Promorar, Itaqui - RS, Brazil, 97650-000}

\affiliation[b]{Instituto de F\'\i sica Te\'orica, Universidade Estadual Paulista, Rua Dr. Bento Teobaldo Ferraz 271 S\~ao Paulo, SP, Brazil, 01140-070 }

\affiliation[c]{Astrophysics Science Division, NASA Goddard Space Flight Center, Greenbelt, MD 20771}

\vspace{-1.5cm}%

\abstract{We consider here three dark matter models with the gauge symmetry of the standard model plus an additional local $U(1)_D$ factor. One model is truly secluded and the other two models begin flipped, but end up secluded. All of these models include one dark fermion and one vector boson that gains mass via the Stueckelberg mechanism.  We show that the would be flipped models provide an example dark matter composed of ``almost least interacting particles" (ALIPs). Such particles are therefore compatible with the constraints obtained from both laboratory measurements and astrophysical observations.}

%\begin{keyword}
%dark matter
%\end{keyword}

\maketitle

\section{Introduction}
\label{sec:int}

The existence of dark matter (DM) requires physics beyond standard model (SM). The so-called dark matter comprises approximately four fifths of the mass of the universe as manifested by its gravitational influence (e.g.,~\cite{Ade:2015xua}). However, at the present time its other fundamental characteristics are still illusive. This is because, aside from its gravitational effects, no other definitive physical evidence of its nature has been found. In particular, severe limits on its interaction with standard model particles have been obtained in the laboratory~\cite{Aprile:2017iyp}--\cite{Fu:2016ega}). This dark matter could well be in hidden sectors comprised of additional particles and forces~\cite{posp08}. Such particles may have either minuscule or nonexistent interactions with SM particles.

Interesting models for physics beyond the SM can be constructed by considering the possibility of an extra local $U(1)_D$ symmetry in addition to the standard SM gauge symmetries.  If the electric charge has a component on the new factor, the model is called ``flipped" otherwise, the model is called ``secluded". In both cases, the vector boson related with the $U(1)_D$ symmetry factor may gain a mass from a spontaneously symmetry breaking through a singlet scalar, and or by the Stueckelberg mechanism~\cite{stue1938} if the vector boson couples with a conserved vector current.

These models present two possibilities. In the secluded model the dark vector does not have a projection on the photon field. Cross sections with SM particles in the laboratory are negligible, but indirect astrophysical signals of dark matter annihilation are potentially observable~\cite{Fortes:2015qka}. On the other hand, in flipped models, the dark vector boson has a component along the photon field and the electric charge operator changes with respect to its definition in the standard model. In principle, this can produce examples of milli-charged dark matter. In both cases, the vector boson related with the $U(1)_D$ symmetry factor may gain a mass from a spontaneously symmetry breaking through a singlet scalar, or by the Stueckelberg mechanism~\cite{stue1938} if the vector boson couples with a conserved vector current.

Here we will compare and discuss the implications of three models with the gauge symmetry of the SM plus an extra local $U(1)_D$ factor with the particle content of the dark sector consisting of a Dirac fermion and a vector field, where the vector field gets mass only by the Stueckelberg mechanism. One model is the secluded model in Ref.~\cite{Fortes:2015qka}. The other models start out as``flipped" models~\cite{Kors:2004dx,Feldman:2007wj} but end up as secluded models after the constraint on the possible neutrino electric charge is taken into account. This is why we will refer to these models as ``putative flipped models".  In the secluded models the dark vector mixes in the kinetics term with the vector boson $B$ related with the $U(1)_Y$ factor of the SM \textit{before} the spontaneous electroweak symmetry breaking (SESB). In the first flipped model there is no kinetic mixing but the vector boson $B$ gets a Stueckelberg mass \textit{after} the SESB mass, besides the mass obtained by the usual Higgs mechanism. This model has been called  Stueckelberg Extension of the Standard Model (StESM). In the second flipped model, a kinetics mixing is added to the StESM.

Unlike in the case of the secluded models, it has been claimed that in the flipped models the dark fermion is millicharged. However we will show here that this is not true because in those models, either the neutrinos gain an electric charge or the electric charges of the known fermions are modified. This result imposes strong constraints on the mixing angles for the neutral vector particles. For all practical purposes, this leaves only a single angle, viz., the electroweak angle $\theta_W$.

We will show that the experimental constraints on the electric charge of the neutrinos and on the measured electric charge of the electron imply that the dark fermion must be almost as neutral as the neutrino. Such bounds do not appear in the secluded model because in that model neutrinos do not gain an electric charge and the charged fermions get the correct charge, $q_igs_W$. Thus in case of the StESM, with or without kinetics mixing, the dark sector decouples from the SM particles. Therefore, in the StESM model the dark matter fermion will not produce any observable effects in either laboratory or astrophysical settings. The StESMs are in fact examples dark matter composed of ``almost least interacting particles" (ALIPs). Such particles are compatible with the constraints obtained from both laboratory measurements and astrophysical observations.

The outline of the paper is as follows: In Section~\ref{sec:secluded} we briefly review the secluded model. We present this model in a different way from that in~\cite{Fortes:2015qka} in order to clarify the difference between this model and the flipped models discussed in Sections \ref{sec:flippedI} and \ref{sec:flippedII}. We consider the StESM without~\cite{Kors:2004dx}, and with the kinetics~\cite{Feldman:2007wj}, in Sections.~\ref{sec:flippedI} and \ref{sec:flippedII} respectively. In Section~\ref{sec:lips}  we show that in fact both StESM models are examples of almost least interacting particles (``ALIPs"). A discussion of the behavior of such particles is given in Section \ref{sec:ss}. Our conclusions appear in Section~\ref{sec:con}.

\section{Secluded Stueckelberg Model}
\label{sec:secluded}

A form of the Stueckelberg model for secluded DM was proposed in Ref.~\cite{Fortes:2015qka}. Here we briefly review rewritten the neutral interactions of model in order to compare this with the flipped models discussed in Secs. III and IV.  We assume a kinetic mixing between a dark vector boson, $V^\prime$, and the $U(1)_Y$ factor of the standard model (SM) with kinetic mixing with the $V^\prime$ occurring \textit{before} spontaneous symmetry breaking being denoted by $B^\prime$.   We will denote $B$ and $V$ the vector fields associated with the $U(1)_D$ factor and the dark vector in the basis with diagonal kinetics terms.  If we start with
\begin{eqnarray}\label{e3}
\mathcal{L^{\prime}}_{kin}&\!=\!& \frac{M_{V}^{2}}{2}V^{\prime}_{\mu}V^{\prime \mu}
\!-\!\frac{1}{4}V^{\prime}_{\mu\,\nu}V^{\prime\mu\nu}+\frac{g_{VB}}{2}V^{\prime}_{\mu\,\nu}B^{\prime\mu\nu}
-\frac{1}{4}B^{\prime}_{\mu\nu}B^{\prime\mu\nu},
\label{k1}
\end{eqnarray}
making the transformation
\begin{equation}
\left(\begin{array}{c}
V^\prime \\B^\prime\end{array}\right)=\frac{1}{\sqrt{1-g^2_{VB}}}\left(
\begin{array}{cc}
1 & 0 \\
g_{VB} & \sqrt{1-g^2_{VB}} \\
\end{array}\right)
\left(\begin{array}{c}
V\\ B
\end{array}
\right),
\label{k2}
\end{equation}
we obtain the diagonal mixing terms:
\begin{eqnarray}
\mathcal{L}_{kin}&=& \frac{1}{2}\bar{M}_{V}^{2}V_{\mu}V^{ \mu}  -\frac{1}{4}V_{\mu\nu}V^{\mu\nu}-\frac{1}{4}B_{\mu\nu}B^{\mu\nu},
\label{k3}
\end{eqnarray}
where $\bar{M}^2_N=M^2_V/\sqrt{1-g^2_{VB}}$.

In the covariant derivatives acting on $H=(0,\,v/\sqrt2)^T$ (unitary gauge) we use $B^\prime$,
\begin{equation}
\mathcal{D}_\mu^H=\mathbf{1}\partial_\mu+i\frac{g}{2}\vec{\tau}\cdot \vec{W}+i\frac{g^\prime}{2}\mathbf{1}B^\prime_\mu,
\label{k4}
\end{equation}
 where $B^\prime$ is given in Eq.~(\ref{k2}). After including the bare term in Eq.~(\ref{k3}) for the field $V$ and the spontaneously symmetry breaking with $v\not=0$,  we obtain the mass matrix in the basis $(V\,B\,W_3)$ of Ref.~\cite{Fortes:2015qka} :
\begin{equation}
M^2_{FPS}=\frac{g^2v^2}{4c^2_W}
\left(
\begin{array}{ccc}
\xi^2 s^2_W+r & \xi s^2_W & -\xi c_Ws_W \\
\xi s^2_W & s^2_W & -c_Ws_W \\
-\xi c_Ws_W & -c_Ws_W & c^2_W
\end{array}
\right),
\label{k5}
\end{equation}
where $c_W\equiv \cos\theta_W, ...$, $\xi=g_{VB}/\sqrt{1-g^2_{VB}}$, $r=M^2_V/M^2_Z$, $M_Z=gv/2c_W$. The matrix in (\ref{k5})  is diagonalized using the matrix
\begin{equation}
\left(
\begin{array}{c}
V\\B\\ W_3 \end{array}
\right)=\left(\begin{array}{ccc}
c_\alpha & s_\alpha & 0\\
s_Ws_\alpha  & -s_Wc_\alpha & c_W\\
-c_Ws_\alpha & c_Wc_\alpha& s_W
\end{array}
\right)
\left(
\begin{array}{c}
Z^\prime \\Z \\ A\end{array}
\right),
\label{k62}
\end{equation}
where $\tan(2\alpha)=2\xi s_W/(1-\xi^2 s^2_W-r)$.

As in Eq.~(\ref{k4}) we have also to use $B^\prime$ in the covariant derivative acting in the fermion sector.
We first consider the lepton sector for both left-handed and right-handed leptons:
\begin{equation}
\mathcal{D}_\mu^{L_L}=\mathbf{1}\partial_\mu+i\frac{g}{2}\vec{\tau}\cdot \vec{W}-i\frac{g^\prime}{2}\mathbf{1}B^\prime_\mu ,\quad
\mathcal{D}_\mu^{l_R}=\mathbf{1}\partial_\mu-i2\frac{g^\prime}{2}\mathbf{1}B^\prime_\mu.
\label{k7}
\end{equation}

As in PDG, we parametrize the neutral currents as follows:

\begin{equation}
\mathcal{L}_{NC}=-\frac{g}{2c_W}\bar{\psi}[\gamma^\mu (g^\psi_V-\gamma_5g^\psi_A)Z_\mu+\gamma^\mu (f^\psi_V-\gamma_5f^\psi_A)Z_\mu]\psi
\label{ncpdg}
\end{equation}
From (\ref{k7}) we get
\begin{equation}
\begin{array}{c}
\mathcal{L}^{leptons}_{NC}=-\frac{g}{2c_W}
[\bar{\nu}_L\gamma^\mu (-W_3c_W+B^\prime s_W)_\mu\nu_L \\ \\
+ \bar{l}_L\gamma^\mu(W_3c_W+B^\prime s_W)_\mu l_L
+2s_W\bar{l}_R\gamma^\mu l_RB^\prime_\mu]
\label{k8}
\end{array}
\end{equation}
\textit{Neutrinos}:

Using the same parametrization as in Eq.~(\ref{ncpdg}), (\ref{k2}) and (\ref{k62}) in (\ref{k8}), we obtain
%\begin{equation}
%\mathcal{L}_\nu=-\frac{g}{2c_W}\bar{\nu}_L\gamma^\mu[(c_\alpha-\xi s_\alpha)Z_\mu-(s_\alpha+\xi c_\alpha)Z^\prime_\mu]\nu_L,
%\label{k9}
%\end{equation}
\begin{equation}
\begin{array}{c}
Q^\nu=0,\quad g^\nu_V=g^\nu_A=\frac{1}{2}(c_\alpha-\xi s_{W} s_\alpha), \\ \\
\quad f^\nu_V=f^\nu_A=\frac{1}{2}(s_\alpha+\xi s_{W} c_\alpha),
\label{ncnus}
\end{array}
\end{equation}
where $Q^\psi$ denote the electric charge of the particle $\psi$. Only when $s_\alpha=0$, neutrinos have the SM interactions with $Z$ i.e. $g^\nu_V=g^\nu_A=1/2$, but even in this case, they interacts with $Z^\prime$ with a strength proportional to $\xi$.  Unlike in other models with kinetic mixing, neutrinos do not couple with the photon.

\textit{Charged Leptons}:

As in the previous case we obtain that their electromagnetic interactions are as those of SM, $e=gs_W=g^\prime c_W$ and the neutral coupling to $Z$ and $Z^\prime$ are
\begin{eqnarray}
&&f^l_V=-\frac{1}{2}\left[(1-4s^2_W)s_\alpha-3\xi s_W c_\alpha\right],f^l_A=-\!\!\frac{1}{2}(s_\alpha+\xi s_W c_\alpha) \nonumber \\  &&
g^l_V=\frac{1}{2}\left[(-1\!\!+\!\!4s^2_W)c_\alpha\!\!-\!\!3\xi s_Ws_\alpha\right],g^l_A\!\!=-\!\!\frac{1}{2}\left(c_\alpha-\xi s_Ws_\alpha\right).
\label{nccl}
\end{eqnarray}
Here $g^l_{V,A}$ coincides with the SM expressions when $s_\alpha=0$, as in \cite{Fortes:2015qka} if we use Eqs.~(A.23) and Eqs.~(A.24) for $g^l_L,g^l_R$, respectively and define $g^l_V=(1/2)(g^l_L+g^l_R)$ and $g^l_A=(1/2)(g^l_L-g^l_R)$.

\textit{Quarks}:

In this sector we have
\begin{equation}
\begin{array}{c}
\mathcal{D}_\mu^{Q_L}=\mathbf{1}\partial_\mu+i\frac{g}{2}\vec{\tau}\cdot \vec{W}+i\frac{1}{3}\frac{g^\prime}{2}\mathbf{1}B^\prime_\mu , \\ \\
\mathcal{D}_\mu^{u_R}=\mathbf{1}\partial_\mu+i\frac{4}{3}\frac{g^\prime}{2}\mathbf{1}B^\prime_\mu, ~~ \mathcal{D}_\mu^{d_R}=\mathbf{1}\partial_\mu-i\frac{2}{3}\frac{g^\prime}{2}\mathbf{1}B^\prime_\mu.
\label{k10}
\end{array}
\end{equation}

From Eqs.~(\ref{k10}), using Eq.~(\ref{k2}) and Eq.~(\ref{k62}),   we obtain
\begin{eqnarray}
&& g^u_V=\frac{1}{2}\left[\left(1-\frac{8}{3}s^2_W\right)c_\alpha+\frac{5}{3}\xi s_Ws_\alpha\right],\; g^u_A=\frac{1}{2}\left(c_\alpha-\xi s_Ws_\alpha\right),\nonumber \\ &&
f^u_V=\frac{1}{2}\left[-\frac{5}{3}\xi s_{W}c_{\alpha}+ s_{\alpha}-\frac{8}{3}s_{W}^{2}c_{\alpha} \right],\; f^u_A=\frac{1}{2}\left[\xi s_{W}c_{\alpha}+s_{\alpha}\right],\nonumber \\&&
g^d_V=\frac{1}{2}\left[\left(-1+\frac{4}{3}s^2_W\right)c_\alpha-\frac{1}{3}\xi s_Ws_\alpha\right],\; g^d_A=-\frac{1}{2}\left(c_\alpha-\xi s_Ws_\alpha\right),\nonumber \\ &&
f^d_V=\frac{1}{2}\left[\frac{1}{3}\left(c_{\alpha}\xi s_{W}+ 4s_{W}^{2}s_{\alpha}\right)-s_{\alpha} \right],\; f^d_A=-\frac{1}{2}\left[c_{\alpha} \xi s_{W} +s_{\alpha} \right].
\label{ncq}
\end{eqnarray}
The angle $\alpha$ is constrained by the neutral current data. However,these constraints are weaker than the astronomical observations that imply $s_\alpha\simeq-4.7\times10^{-7}$~\cite{Fortes:2015qka}.

\textit{The Dark Fermion}:

  Finally the neutral interactions of the dark fermion $\eta$, which has couplings as $g_\eta \bar{\eta}\gamma^\mu\eta V_{\mu}$ and we have:
\begin{equation}
\begin{array}{c}
Q^\eta = 0,\;\;\;
g^\eta_V = \frac{g_\eta}{
\sqrt{1-g^2_{VB} }} s_\alpha, \;\;\; g^\eta_A=0, \\ \\
f^\eta_V = \frac{g_\eta }{
\sqrt{1-g^2_{VB} }}c_\alpha,\;\;\;f^\eta_A = 0.
\label{eta1}
\end{array}
\end{equation}
In this model the dark fermion and neutrino do not carry electric charge and that the couplings of the dark fermion with $Z$ and $Z^\prime$ are vectorial only. The model has a free angle, $\alpha$, and is possible to constraint it as it was done in Ref.~\cite{Fortes:2015qka}.

\section{Flipped Stueckelberg model I}
\label{sec:flippedI}

In this model a dark fermion and dark Stueckelberg vector boson coupled with a conserved vector current are introduced, but this time without  kinetic mixing. Moreover, a Stueckelberg mass for the vector boson of SM $U(1)_Y$ factor is added. Since there is no kinetic mixing, the field $B$ in the covariant derivative is already the field coupled with the SM fields. This field gains a Stueckelberg bare mass term. This model has been dubbed Stueckelberg extension of the standard model (StESM), and it was proposed in Ref.~\cite{Kors:2004dx}.

When $\xi=0$ (no kinetic mixing) in Eq.~(\ref{k4}), and introducing a bare mass $M_B$ for the field $B$, we have the model of Ref.~\cite{Kors:2004dx} in which the mass matrix of the neutral vector bosons is given by
\begin{equation}
M^2_{KN}=\frac{g^2v^2}{4c^2_W}
\left(
\begin{array}{ccc}
r^2_1 & r_1r_2  & 0  \\
r_1r_2 & r^2_2+s^2_W & -c_Ws_W \\
0 & -c_Ws_W & c^2_W
\end{array}
\right)
\label{knmass}
\end{equation}
where $r_1=M_V/M_Z$, $r_2=M_B/M_Z$ [$M_B(M_V)$] was denoted by $M_2(M_1)$ in Ref.~\cite{Kors:2004dx}).

We assume that the known fermions are not charged under the dark $U(1)_D$, hence the covariant derivatives are the usual ones and the mixing with the dark sector is only through the mass matrix involving $V,B,W_3$. The matrix in (\ref{knmass}) is diagonalized by the following orthogonal matrix, {\bf O}:
\begin{equation}
{\bf O}=\left(
%\begin{array}{c}
%V\\B\\ W_3 \end{array}
%\right)=\left
\begin{array}{ccc}
c_\psi c_\phi-s_\theta s_\phi s_\psi  & -s_\psi c_\phi-s_\theta s_\phi c_\psi &  -c_\theta s_\phi\\
c_\psi s_\phi+s_\theta c_\phi s_\psi & -s_\psi s_\phi+s_\theta c_\phi c_\psi  & c_\theta c_\phi \\
-c_\theta s_\psi & -c_\theta c_\psi  & s_\theta
\end{array}
\right),
\label{kn1}
\end{equation}
{\rm so that}
\begin{equation}
\left(
\begin{array}{c}
V\\B\\ W_3
\end{array}\right)
= {\bf O} \left(\begin{array}{c}
Z^\prime\\ Z \\ A
\end{array}
\right),
\end{equation}
where
\begin{equation}
\tan\phi=\frac{M_B}{M_V},\quad \tan\psi=\frac{\tan\theta_W M^2_W}{c_W[M^2_{Z^\prime}(1-\tan^2\theta_W)M^2_W ]}\tan\phi.
\label{defkn}
\end{equation}
One of the angles in the matrix in (\ref{kn1}) has to be, within the experimental error, equal to the weak mixing angle, $\theta_W$.

Unlike the matrix in equation (\ref{k62}), in this case, $V$ has a component along the photon field. This is a flipped model. The dark fermion and neutrinos will get an electric charge and the known charged fermion will also get a small non-standard electric charge. Explicitly
\begin{eqnarray}
&& Q^\nu =-\frac{1}{2}gs_W(-1+ c_\phi), \;\;\; Q^\eta= -g_\eta c_W s_\phi ,\nonumber \\ &&
Q^l_L= -\frac{1}{2}gs_W(1+c_\phi),\;\; Q^l_R=-gs_W c_\phi ,\nonumber \\&&
Q^u_L=\frac{1}{2}gs_W\left(1+\frac{1}{3}c_\phi\right), \;\; Q^u_R=\frac{2}{3}gs_Wc_\phi, \nonumber \\&&
Q^d_L=-\frac{1}{2}gs_W\left(1-\frac{1}{3}c_\phi\right), \;\; Q^d_R=-\frac{1}{3}gs_Wc_\phi,
\label{knec1}
\end{eqnarray}
where $gs_W=\vert e\vert$.
Notice that only if $\phi=0$, $Q^l_L=Q^l_R=-gs_W$ and $Q^\nu=0$.  Similarly for quarks, only when $\phi=0$ we obtain the usual electric charge  for the known charged fermions.
On the other hand, astrophysical constraints on neutrino electric charge imply $Q^\nu<10^{-19}\vert e\vert$~\cite{Studenikin:2012vi}. Stronger constraints come from $\beta$-decay: $Q^\nu<10^{-21}\vert e \vert$~\cite{Marinelli:1983nd,Baumann:1988ue}. In fact, in order to $Q^\nu$ be compatible with the experimental upper limit on a possible neutrino electric charge then $(1-c_\phi)<2\times 10^{-21}$. This implies that $\phi
\approx 0$ and $M_B \approx 0$. In accordance with the definition in Eq.~(\ref{defkn}), the angle $\psi \approx 0$ is also practically zero. Hence, the model has only one angle: $\theta_W$.

We can  calculate the neutral current couplings in the fermion sector using still all the angles in (\ref{kn1}). \\

\newpage

\textit{Neutrinos}:

\begin{eqnarray}
&& g^\nu_V =g^\nu_A=\frac{1}{2}[(c_{W}^{2}+s^2_Wc_\psi)c_\phi-s_{W}s_{\psi}s_{\phi}], \nonumber \\&&
f^\nu_V=f^\nu_A =\frac{1}{2}\left[ (s_{W}^{2}c_{\phi}+c_{W}^{2})s_{\psi}+s_{W}s_{\psi}c_{\phi} \right].
\label{knnc1}
\end{eqnarray}

\textit{Charged Leptons}:

\begin{eqnarray}
&& g^l_V =\frac{1}{2}[(1-3s^2_{W}c_\phi-s^2_{W})c_\psi+3s_Ws_{\phi}s_{\psi})], \nonumber \\&&
 g^l_A=\frac{1}{2}[(c^2_{W}+s^2_Wc_\phi)c_\psi -s_Ws_\phi s_\psi ], \nonumber  \\ &&
f^l_V= \frac{1}{2}\left[\left( c^2_W-3s^2_Wc_\phi\right)s_\psi-3s_W c_\psi s_\phi \right],\nonumber \\&&
f^l_A =\frac{1}{2}\left[ \left( c^2_W+s^2_Wc_\phi\right)s_\psi+s_W c_\psi s_\phi\right].
\label{knnc2}
\end{eqnarray}

\textit{$Up$-quarks}:

\begin{eqnarray}
&& g^u_V =\frac{1}{2}\left[\left(c^2_W-\frac{5}{3}s^2_Wc_\phi \right)c_\psi+\frac{5}{3}s_Ws_\phi s_\psi\right], \nonumber \\ &&
g^u_A=\frac{1}{2}\left[(c^2_W+s^2_Wc_\phi)c_\psi-s_Ws_\psi s_\psi  \right], \nonumber \\ &&
f^u_V= \frac{1}{2}\left[\left(c^2_W-\frac{5}{3}s^2_Wc_\phi \right)s_\psi-\frac{5}{3}s_W s_\phi c_\psi\right],\nonumber \\ &&
f^u_A =\frac{1}{2}\left[ (c^2_W+s^2_Wc_\phi)s_\psi+ s_W s_\phi c_\psi\right]
\label{knnc3}
\end{eqnarray}

\textit{$Down$-quarks}:

\begin{eqnarray}
&& g^d_V =\frac{1}{2}\left[\left( -c^2_W+\frac{1}{3}s^2_Wc_\phi\right)c_\psi-\frac{1}{3}s_Ws_\phi s_\psi \right], \nonumber \\&&
g^d_A=-\frac{1}{2}\left[ \left(c^2_W+s^2_Wc_\phi \right)c_\psi-s_W s_\phi s_\psi \right], \nonumber \\ &&
f^d_V =-\frac{1}{2}\left[ \left(c^2_W+s^2_Wc_\phi \right)s_\psi+s_Ws_\phi c_\psi \right], \nonumber \\&& 
f^d_A =-\frac{1}{2}\left[ \left(c^2_W+s^2_Wc_\phi \right)s_\psi+s_Ws_\phi c_\psi \right].
\label{knnc4}
\end{eqnarray}

\textit{The Dark Fermion}:

\begin{eqnarray}
&& g^\eta_V=-g_\eta(s_\psi c_\phi+s_W s_\phi c_\psi),\;\; g^\eta_A=0,\nonumber \\ && f^\eta_V=g_\eta(c_\psi c_\phi-s_W s_\phi s_\psi),\;\; f^\eta_A=0.
\label{eta2}
\end{eqnarray}

Notice that the dark fermion has only vector interactions. However, if we now take into account the neutrino electric charge constraint discussed above, $\phi = 0$ in order for $Q^\nu$ in equation (\ref{knec1}) to be compatible with the upper experimental bound limits. Because of the relations in (\ref{defkn}), $ \psi= 0$ as well. In this situation
all of the coupling constants $g_V,g_A$ thus reduce to their SM values. Therefore the dark fermion couples only with $Z^\prime$ and $Q^\eta = 0$. Hence there is no millicharged dark matter in the model of Ref.~\cite{Kors:2004dx}. The $B$ vector of the standard model cannot receive (if any) a  considerable Stueckelberg bare mass.

Above we assumed that $\theta_W = \theta$, another possibility used in Ref.~\cite{Kors:2004dx} is $\tan\theta= \theta_W c_\phi$. Below we show that, in fact, $\tan\theta= \tan\theta_W c_\phi$ implies $\tan\theta= \tan\theta_W $.

Let us again consider the electric charge of the known leptons in the flipped models of Sec.~\ref{sec:flippedI} and \ref{sec:flippedII}. As in the SM, the interaction of leptons with the vector $W_3$ and $B$ fields are given by
\begin{equation}
\mathcal{L}_l=-\frac{g}{2}\left[\bar{\nu}_L(W_3-\tan\theta_WB)\bar{\nu}_L +\bar{l}_L(-W_3-\tan\theta_WB)\bar{l}_L\right]+g_Y\bar{l}_R (B)l_R,
\label{knn}
\end{equation}	
where $g_Y$ is the gauge coupling of the $U(1)_Y$ factor of the SM. From (\ref{kn1}) we obtain
\begin{eqnarray}
&& W_3=-c_\theta s_\psi Z^\prime -c_\theta c_\psi Z +s_\theta A,\nonumber \\ &&
B=(c_\psi s_\phi +s_\theta c_\phi s_\psi )Z^\prime +(-s_\psi s_\phi+s_\theta c_\phi c_\psi)Z+c_\theta c_\phi A .
\label{kn11}
\end{eqnarray}
Using equations (\ref{kn11}) and (\ref{knn}) we find that the electric charge of the leptons are given by
\begin{eqnarray}
&& Q^\nu=-\frac{g}{2c_W}\cdot c_W c_\theta (\tan\theta-\tan\theta_W c_\phi), \nonumber \\ &&
Q^l_L=\frac{g}{2c_W}\cdot c_W c_\theta (\tan\theta+\tan\theta_W c_\phi),\quad Q^l_R=g \tan\theta_W c_\theta c_\phi,
\label{kn2n}
\end{eqnarray}

Here we have two possibilities.
\begin{enumerate}
\item[a)] $\theta=\theta_W$ 	
and in this case we have in Eq.~(\ref{kn2n})
\begin{eqnarray}
&& Q^\nu=-\frac{gs_W}{2} (1- c_\phi), \nonumber \\ &&
Q^l_L=\frac{g}{2}s_W(1+c_\phi),\quad Q^l_R=g s_Wc_\phi,
\label{kn5}
\end{eqnarray}
as in Eq.~(\ref{knec1}). Hence, as we said $\phi$ is for all practical purposes zero.
	
Or we can choose
\item[b)] $\tan\theta=\tan\theta_W c_\phi$. %In this case the electric charge of neutrino vanishes, $Q^\nu=0$.
However, in case b) we have two possibilities: doing $\tan\theta_Wc_\phi\to \tan\theta$ in Eq.~(\ref{kn2n}), we obtain
\begin{equation}
Q^\nu=0,\quad Q^l_L \equiv  Q^l_R=g s_\theta,
\label{kn6}
\end{equation}
or, doing in (\ref{kn2n})  $\tan\theta\to \tan\theta_Wc_\phi$
%Nevertheless, notice that in case b) we can make in (\ref{kn2n}) that from (\ref{kn6}), that $\theta\equiv \theta_W$. It %means again that $c_\phi=0$ .
we obtain %On the other hand, if  had choosen, then
\begin{equation}
Q^\nu=0,\quad Q^l_L= Q^l_R=gc_\theta \tan\theta_Wc_\phi.
\label{kn7}
\end{equation}

\end{enumerate}

\section{Flipped Stueckelberg model II}
\label{sec:flippedII}

It is possible to add a kinetic mixing term as the $g_{VB}$ term in Eq.~(\ref{k1}) to the flipped model I that was considered in Sec.~\ref{sec:flippedI}. This was done in Ref.~\cite{Feldman:2007wj}, where the coupling constant $g_{VB}$ was denoted by $\delta$. In principle, such models have two parameters in the dark vector sector: a kinetic mixing, as in the secluded model in Sec.~\ref{sec:secluded}, and a bare Stueckelberg mass $M_B$ as in the flipped I model in Sec.~\ref{sec:flippedI}.

In this case the orthogonal matrix in Eq.~(\ref{kn1}) is still the same and all the results in Eqs.~(\ref{knec1}) -- (\ref{eta2}) are valid. Hence all the results in Sec.~\ref{sec:flippedI} are still valid.  The neutrino and the dark fermion $\eta$ would have an electric charge as in Eqs.~(\ref{knec1}). All of the couplings in Eqs.~(\ref{knnc1}) -- (\ref{eta2}) are the same as well.

In this model, as in the previous model, the experimental constraints of a possible neutrino electric charge imply that $\phi\approx0$. However, the definition in Eq.~(\ref{defkn}) is replaced by~\cite{Feldman:2007wj}:
\begin{equation}
\tan\phi=\bar{\epsilon},\;\;\; \tan2\psi=\frac{2s_WM^2_Z\tan\phi}{M^2_V-M^2_Z+(M^2_V-M^2_Z-M^2_W)\bar{\epsilon}^2}
\label{flndef}
\end{equation}
where $\bar{\epsilon}=r_1-\delta$ (this is the parameter in the kinetics mixing in the notation of Ref.~\cite{Feldman:2007wj}) and as in the case of the model in the previous section, when $\phi=0$, then $\psi=0$ and the dark sector decouple from the SM particles.
Thus, this model actually becomes secluded and is similar to that in Sec.~\ref{sec:secluded}. Thus, there is no millicharged dark matter in this case. However, in this case  $\tan\phi=0$ means, according to Eq. (\ref{flndef}) that $\epsilon=\delta$ and then it is possible to have a  bare mass for the vector boson $B$ but the dark sector decouple from the SM particles.

Notice that (\ref{kn6}) and (\ref{kn7}) are compatible with each other if, and only if, $\theta=\theta_W$ and $c_\phi=0$ (or extremely small). Hence, we can from the very beginning use $\theta=\theta_W$ as we have done in Secs.~\ref{sec:flippedI} and \ref{sec:flippedII}.

\section{LIPs and ALIPs}
\label{sec:lips}

In the dark sector we can distinguish two cases of DM models that are consistent with the observational astrophysical and laboratory constraints: (1) particles that interact 
%among themselves and 
with the particles of the standard model only through gravitational interactions, labeled Least Interacting Particles (LIPs)~\cite{Matsas:1998zm}, and (2) particles which we call Almost LIPs (ALIPs). ALIPs interact with the particles of the standard model only by gravitational interactions, but also interact with each other through new interactions and which may have their own symmetry (see Table 1).

A true LIP particle is one that interacts only gravitationally. Examples of true LIPs are primordial black holes. Such particles, in order to make up the dark matter, must be stable against Hawking radiation~\cite{ha71} up until the present epoch. Thus their mass must be greater than $10^{15}$ g. Astrophysical constraints on such primordial black holes are discussed in Ref.~\cite{ca10} and~\cite{cl17}.

In the models considered in Secs.~\ref{sec:flippedI} and \ref{sec:flippedII}, the dark sector consists of a massive vector boson $Z^\prime$ and a Dirac fermion $\eta$ that couples to the standard model particles mainly via gravitation.
In such models the fields in the dark sector can explain the matter relic density but no direct effects can be observed in astrophysical processes or in laboratory.
The Kors and Nath model~\cite{Kors:2004dx} is an ALIP model, since the angle $\phi$ is not exactly zero but rather is a small number, taken to be compatible with the upper bound of a possible neutrino electric charge. This model has no angles besides the weak angle, $\theta_W$, in the visible sector.

Let us assume only that $m_V > m_\eta$. In the early universe $Z^\prime+Z^\prime\leftrightarrow \eta+\bar{\eta}$, and $Z^\prime$ and $\eta$ were in equilibrium. Then at a given later time, say $t_{0}$, only $V+V\to \eta+\bar{\eta}$ could occur and the fermion $\eta$ decoupled. The rate of this reaction depends on the coupling constant $g_\eta$ and the masses of the fermion and vector bosons. When $\Gamma^{-1}>H^{-1}(t_{0})$, $\eta$ decoupled and its number remained constant thereafter, thus accounting for the 27\% of cold dark matter (CDM) observed by Planck Collaboration at present~\cite{Ade:2015xua}.

Denoting $\rho_{_{crit}}$ the critical density of the universe
\begin{equation}
\Omega_{_{CDM}}=\frac{\rho_{_{CDM}}}{\rho_{crit}}=0.258,
\label{dm1}
\end{equation}
being %[also  PDG p. 120 (but I approximated to 3 decimal places)] {\bf Put in PDG reference.}
\begin{equation}
\rho_{_{crit}}=1.054\times10^{-5}\,h^2\,\frac{\textrm{GeV}}{\textrm{cm}^3}=0.48\times10^{-5}\,\frac{\textrm{GeV}}{\textrm{cm}^3},
\label{dm2}
\end{equation}
where we have used $h=0.678$~\cite{Olive:2016xmw}.

From (\ref{dm1}) and (\ref{dm2}) we obtain
\begin{eqnarray}
\rho_{_{CDM}}&=&1.24\times10^{-6}\,\frac{\textrm{GeV}}{\textrm{cm}^3}=
1.24\,\frac{\textrm{GeV}}{\textrm{m}^3},\nonumber \\&=&
2.22\times10^{-24}\frac{\textrm{g}}{\textrm{m}^3}.
%1.78\times 10^{-9}\frac{g}{\textrm{km}^3}.
\label{dm3}
\end{eqnarray}

This shows that an ALIP with mass of 1.24 GeV per cubic meter can explain the observed relic density. The ALIP mass is constrained by the unitary limit for a Dirac fermion
\begin{equation}
\Omega_{tot} h^2  \sim 0.5 \ge 8.7 \times 10^{-6} [m_{\chi}({\rm TeV})]^2 ,
\end{equation}
which gives an upper limit on an ALIP Dirac fermion mass  of  $\sim 240 $ TeV~\cite{Griest:1989wd}. This is in contrast to LIP particles that can much larger masses.

It is important to notice the difference between the model in Sec.~\ref{sec:secluded} and those in Sec.~\ref{sec:flippedII}. In both there is a mixing in the vector kinetic terms, but as in the former, the kinetic terms are diagonalized \textit{before} the spontaneous symmetry breaking (SSB). On the other hand, in the model of Sec.~\ref{sec:flippedII}~\cite{Kors:2004dx,Feldman:2007wj}, the diagonalization of the kinetic mixing is done
\textit{after} the SSB, $M^2_{FLN}=K^TM^2_{KN}K$ where $K$ is the $3\times3$ extension of the matrix defined in Eq.~(\ref{k2}) and $M^2_{KN}$ is the matrix in Eq.~(\ref{knmass}). The result is the matrix of Eq.~(A2) of Ref.~\cite{Feldman:2007wj} that has determinant zero.
In fact, if in the model of Sec.~\ref{sec:flippedII} the kinetics mixing had occurred \textit{before} the SSM, the mass matrix of the neutral vector bosons would be
\begin{equation}
M^2_{FLN}=\frac{g^2v^2}{4c^2_W}
\left(
\begin{array}{ccc}
\xi^2 s^2_W+r^2_1 & \xi s^2_W+r_1r_2 & -\xi c_Ws_W \\
\xi s^2_W+ r_1r_2& s^2_W+r^2_2 & -c_Ws_W \\
-\xi c_Ws_W & -c_Ws_W & c^2_W
\end{array}
\right),
\label{k6}
\end{equation}
that has a determinant equal to $-c^2_W(-4+r_1)r_1r^2_2$ and the photon acquires a nonzero mass.

\section {Small Scales}
\label{sec:ss}

In this section we will give a brief preview about the model behavior under small scales.  We have analysed the self-interacting dark matter case. For this calculation we have considered the process $\eta+\eta\rightarrow \eta + \eta$. In principle this calculation is valid for the Flipped Stueckelberg model I and II since we leave $g_{\eta}$ as a generic coupling.

With this process we can have a good estimative of the other related  processes like $\eta+\bar{\eta}\rightarrow \eta+\bar{\eta}$ since in this case we have the crossing symmetry.

 Lets consider now the process $\eta+\eta\rightarrow \eta + \eta$. We had taken into account the statistical factor for identical particles and made the substitution for $s=4m_{\eta}^2+ m_{\eta}^2 v^2$ and we had taken the limit of $v\rightarrow 0$.
In the Eq. (\ref{SI}) we express $\sigma/m$. This number must be in the
0.5 cm$^2$/g $< \sigma/m <$ 1 cm$^2$/g range in order to solve the core-cusp and the too-big-to-fail problem on small scales. These limits are in agreement with the other astrophysical constraints on larger scales.
\begin{eqnarray}
  \frac{\sigma}{m} &=& \frac{g_{\eta}^4 m_{\eta}}{16\pi M_{ZP}^{4}} .
  \label{SI}
\end{eqnarray}
So, for these type of model we should use Eq. (\ref{SI}) to give a estimative of the cross-section over the mass.

Now we consider the dwarf galaxy problem. For the central region of a typical dwarf galaxy according to Ref.~\cite{Oh} we should have $\rho_{DM} \sim 0.1 M_{\odot}$/pc$^{3}$ and $v\sim 50$ km/s \cite{Tulin}. If we convert this constraint to units of g/m$^{3}$ and compare this number with the mass density obtained in equation ~(\ref{dm3}) for our ALIP model, we find the required $\rho_{DM}\sim 6.73\times 10^{-18}$ g/m$^{3}$ is six orders of magnitude larger than the number obtained in equation ~(\ref{dm3}). Thus, this model cannot explain the missing satellite problem.

\section{Conclusions}
\label{sec:con}

Of the three models that we have considered here, only the secluded model of Sec.~\ref{sec:secluded}, i.e., in Ref.~\cite{Fortes:2015qka}, may induce effects that can be observed by astronomers or in laboratory. However, recent experiments trying to observe direct effect of WIMPS, viz., XENON1T~\cite{Aprile:2017iyp}, LUX \cite{Akerib:2016vxi}, and PANDA-II~\cite{Fu:2016ega}, have not obtained any evidence for dark matter particles, with the upper limits on the spin-independent cross section reaching values as low as $\sim 10^{-46}$ cm$^2$. \cite{Ablikim:2017aab}

Astrophysical searches for dark matter (DM) annihilation into $\gamma$-rays have also been somewhat disappointing. While the $\gamma$-ray excess in the galactic center region (GCE) can have a DM annihilation interpretation (e.g.,~\cite{Fortes:2015qka, abaz2016}), such an interpretation appears to be somewhat in conflict with upper limits from dwarf spheroidal satellite galaxies, at least if the main annihilation models are through the $t\bar{t}$ and $\tau \bar{\tau}$ channels ~\cite{Albert2017}. A strong alternate interpretation of the GCE is that it is made up of emission from point sources such as pulsars~\cite{bart2016,TheFermi-LAT:2017vmf}.

Given these present negative empirical results, the possible existence of dark matter particles that mainly, or possibly only interact gravitationally with SM particles (ALIPs or LIPs) is an interesting alternative theoretical option to explore, as we have considered here. 

\section*{Acknowledments}
VP would like to thank CNPq for partial support and is also  thankful for the support of FAPESP funding Grant No. 2014/19164-6 and to G. E. A. Matsas for useful discussions.

\newpage

\vskip .5cm
%\begin{center}
\begin{table}[ht]
	\begin{tabular}{||l|ccccc||} \hline
		Particle/Interactions& Gravitational & weak & electromagnetic & strong &  dark interactions \\ \hline
		Quarks & +  & + & + & + &-  \\ \hline
		Charged leptons & + & + & +  & - & -   \\ \hline
		Neutrinos & + & + & - & - & -\\ \hline
		ALIPS &+ &- & - & - &+ \\ \hline
		LIPS &+ & - & - & -  &- \\ \hline
	\end{tabular}
	\protect\caption{Particle types and their interactions.}
	\label{t26}
\end{table}


\newpage

\begin{thebibliography}{99}
%\cite{Ade:2015xua}
\bibitem{Ade:2015xua}
  P.~A.~R.~Ade {\it et al.} [Planck Collaboration],
%\textrm{Planck 2015 results. XIII. Cosmological parameters},
% Astron.\ Astrophys.\  {\bf 594}, A13 (2016);
 %doi:10.1051/0004-6361/201525830
%  [arXiv:1502.01589 [astro-ph.CO]].

\bibitem{Aprile:2017iyp}
E.~Aprile {\it et al.} [XENON Collaboration], [arXiv:1705.06655] [astro-ph.CO].

%\cite{Aprile:2016swn}
\bibitem{Aprile:2016swn}
E.~Aprile {\it et al.} [XENON100 Collaboration],
%\textrm{XENON100 Dark Matter Results from a Combination of 477 Live Days},
 Phys.\ Rev.\ D {\bf 94}, no. 12, 122001 (2016)
 % doi:10.1103/PhysRevD.94.122001

%\cite{Akerib:2016vxi}
\bibitem{Akerib:2016vxi}
D.~S.~Akerib {\it et al.} [LUX Collaboration],
%\textrm{Results from a search for dark matter in the complete LUX exposure},
  Phys.\ Rev.\ Lett.\  {\bf 118}, no. 2, 021303 (2017).
  %doi:10.1103/PhysRevLett.118.021303

%\cite{Fu:2016ega}
\bibitem{Fu:2016ega}
C.~Fu {\it et al.} [PandaX-II Collaboration],
%\textrm{Spin-Dependent Weakly-Interacting-Massive-Particle--Nucleon Cross Section Limits from First Data of PandaX-II Experiment},
Phys.\ Rev.\ Lett.\  {\bf 118}, no. 7, 071301 (2017).
%doi:10.1103/PhysRevLett.118.071301

\bibitem{posp08}
M.~Pospelov, A. Ritz, and M. Voloshin,
 % \textrm{Secluded WIMP dark matter},
  Physics Letters B {\bf 662}, 53 (2008).

\bibitem{stue1938}
%\textrm{Wechselwirkungskr\"{a}fte in der Elektrodynamik und in der Feldtheorie der Kr\"{a}fte},
E. C. G. Stueckelberg, Helv. Phys. Acta 11, 225 (1938).

%\cite{Fortes:2015qka}
\bibitem{Fortes:2015qka}
E.~C.~F.~S.~Fortes, V.~Pleitez and F.~W.~Stecker,
%\textrm{Secluded WIMPs, dark QED with massive photons, and the galactic center gamma-ray excess},
Astropart.\ Phys.\  {\bf 74}, 87 (2016).
%  doi:10.1016/j.astropartphys.2015.10.010
%[arXiv:1503.08220 [hep-ph]].

%\cite{Kors:2004dx}
\bibitem{Kors:2004dx}
B.~Kors and P.~Nath,
%\textrm{A Stueckelberg extension of the standard model},
Phys.\ Lett.\ B {\bf 586}, 366 (2004).
%doi:10.1016/j.physletb.2004.02.051
%[hep-ph/0402047].

%\cite{Feldman:2007wj}
\bibitem{Feldman:2007wj}
D.~Feldman, Z.~Liu and P.~Nath,
%\textrm{The Stueckelberg Z-prime Extension with Kinetic Mixing and Milli-Charged Dark Matter From the Hidden Sector},
Phys.\ Rev.\ D {\bf 75}, 115001 (2007).
%  doi:10.1103/PhysRevD.75.115001
%[hep-ph/0702123 [HEP-PH]].

%\cite{Studenikin:2012vi}
\bibitem{Studenikin:2012vi}
A.~I.~Studenikin and I.~Tokarev,
%\textrm{Millicharged neutrino with anomalous magnetic moment in rotating magnetized matter},
Nucl.\ Phys.\ B {\bf 884}, 396 (2014).
%doi:10.1016/j.nuclphysb.2014.04.026
%[arXiv:1209.3245 [hep-ph]].

%\cite{Marinelli:1983nd}
\bibitem{Marinelli:1983nd}
M.~Marinelli and G.~Morpurgo,
%\textrm{The Electric Neutrality of Matter: A Summary},
Phys.\ Lett.\ B {\bf 137}, 439 (1984).
%doi:10.1016/0370-2693(84)91752-0

%\cite{Baumann:1988ue}
\bibitem{Baumann:1988ue}
J.~Baumann, J.~Kalus, R.~Gahler and W.~Mampe,
%\textrm{Experimental Limit for the Charge of the Free Neutron},
Phys.\ Rev.\ D {\bf 37}, 3107 (1988).
%doi:10.1103/PhysRevD.37.3107

%\cite{Davidson:2000hf}
\bibitem{Davidson:2000hf}
S.~Davidson, S.~Hannestad and G.~Raffelt,
%\textrm{Updated bounds on millicharged particles},
JHEP {\bf 0005}, 003 (2000).
%doi:10.1088/1126-6708/2000/05/003
%[hep-ph/0001179].

%\cite{Matsas:1998zm}
\bibitem{Matsas:1998zm}
  G.~E.~A.~Matsas, J.~C.~Montero, V.~Pleitez and D.~A.~T.~Vanzella,
%\textrm{Dark matter: The Top of the iceberg?}
Contributions to \textsl{Topics in Theoretical Physics: Festschrift for A. H. Zimerman}, Instituto de F\'\i sica Te\'orica/UNESP- S\~ao Paulo, 1998, Edited by H. Aratyn, L. A. Ferreira and J. F. Gomes, p. 219; hep-ph/9810456 (1998).

\bibitem{ha71}
S. Hawking, Mon. Not. R. astr. Soc. {\b 152} 75 (1971).

\bibitem{ca10}
B. J. Carr, Phys. Rev. D 81, 104019 (2010).

\bibitem{cl17}
S. J. Clark et al. Phys. Rev. D 95, 083006 (2017).

%\cite{Olive:2016xmw}
\bibitem{Olive:2016xmw}
C.~Patrignani {\it et al.} [Particle Data Group],
%\textrm{Review of Particle Physics},
Chin.\ Phys.\ C {\bf 40}, no. 10, 100001 (2016).
%%doi:10.1088/1674-1137/40/10/100001

%\cite{Griest:1989wd}
\bibitem{Griest:1989wd}
K.~Griest and M.~Kamionkowski,
%\textrm{Unitarity Limits on the Mass and Radius of Dark Matter Particles},
Phys.\ Rev.\ Lett.\  {\bf 64}, 615 (1990).
%doi:10.1103/PhysRevLett.64.615

%\cite{Ablikim:2017aab}
\bibitem{Ablikim:2017aab}
M.~Ablikim {\it et al.} [BESIII Collaboration],
%``Dark Photon Search in the Mass Range Between 1.5 and 3.4 GeV/$c^2$,''
Phys.\ Lett.\ B {\bf 774}, 252 (2017).
%doi:10.1016/j.physletb.2017.09.067


\bibitem{abaz2016}
%@ARTICLE{2016PhRvD..93h3514A,
   {{Abazajian}, K.~N. and {Keeley}, R.~E.},
%    \textrm{Bright gamma-ray Galactic Center excess and dark dwarfs: Strong tension for dark matter annihilation despite Milky Way halo profile and diffuse emission uncertainties},
Phys.\ Rev.\ D  {\bf 93}, 083514, (2016),
%archivePrefix = "arXiv",
 %  [arXiv:1510.06424][hep-ph].
% primaryClass = "hep-ph",

\bibitem{Oh}
  S.~H.~Oh, W.~J.~G.~de Blok, E.~Brinks, F.~Walter and R.~C.~Kennicutt, Jr,
  %``Dark and luminous matter in THINGS dwarf galaxies,''
  Astron.\ J.\  {\bf 141}, 193 (2011)
  doi:10.1088/0004-6256/141/6/193
  [arXiv:1011.0899 [astro-ph.CO]].

  \bibitem{Tulin}
  S.~Tulin and H.~B.~Yu,
  %``Dark Matter Self-interactions and Small Scale Structure,''
  arXiv:1705.02358 [hep-ph].
 
\bibitem{Albert2017}
%@ARTICLE{2017ApJ...834..110A,
   {{Albert}, A. et al. %and {Anderson}, B. and {Bechtol}, K. and {Drlica-Wagner}, A. and
	%{Meyer}, M. and {S{\'a}
	Fermi-LAT and Collaborations}, % M. and {Strigari}, L. and
%	  \textrm{Searching for Dark Matter Annihilation in Recently Discovered Milky Way Satellites with Fermi-LAT},
Astrophys. J., {\bf 834}, 110 (2017), [arXiv:1611.03184][astro-ph.HE].

\bibitem{bart2016}
{{Bartels}, R. and {Krishnamurthy}, S. and {Weniger}, C.},
%\textrm{Strong Support for the Millisecond Pulsar Origin of the Galactic Center GeV Excess},
{Physical Review Letters} {\bf 116}, 051102 (2016).

%\cite{TheFermi-LAT:2017vmf}
\bibitem{TheFermi-LAT:2017vmf}
M.~Ackermann {\it et al.} [Fermi-LAT Collaboration],
\textrm{The Fermi Galactic Center GeV Excess and Implications for Dark Matter},
Astrophys.\ J.\  {\bf 840}, no. 1, 43 (2017)
%doi:10.3847/1538-4357/aa6cab

\end{thebibliography}
\end{document}